\documentclass[useAMS,usenatbib,usegraphicx]{mn2e}

\title[The outer wind of $\gamma$ Velorum]{The outer wind of $\gamma$ Velorum}
\author[P. F. Roche, M. D. Colling and M.J. Barlow ]{  P. F. Roche$^{1}$\thanks{E-mail: p.roche1@physics.ox.ac.uk (PFR)}, M. D. Colling$^{1}$ and M.J. Barlow$^2$ \\
$^{1}$Astrophysics, Department of Physics, University of Oxford, DWB, Keble Road, Oxford OX1 3RH \\
$^2$Dept Physics and Astronomy, University College London, Gower Street, London WC1E 6BT}
\begin{document}

\date{Accepted 2012 December 00. Received 2012 July  00; in original form 2011 November 00}

\pagerange{\pageref{firstpage}--\pageref{lastpage}} \pubyear{2011}

\maketitle

\label{firstpage}

\begin{abstract}
Fine-structure mid-infrared emission lines with critical densities in the regime $10^{4} \le $  n$_{crit}$  $\le 10^{6}$ cm$^{-3}$  can be employed to probe the outflow from Wolf-Rayet stars  at radii of $\sim10^{15}$ cm.  
Narrow-band mid-infrared imaging and spectroscopy of the nearest WR star to the sun, $\gamma$ Velorum is analysed for spatially resolved forbidden line emission in the Wolf-Rayet outer wind. The [{S\,\sc iv}] 10.52-$\mu$m and [{Ne\,\sc ii}] 12.81-$\mu$m emission regions are found to be spatially extended, compared to unresolved continuum and He and C recombination line emission. The [{S\,\sc iv}] and [{Ne\,\sc ii}] emission line distributions have a high degree of  azimuthal symmetry, indicating a spherically symmetric outflow.   A model wind with a modest degree of clumping (clumping factor $f\sim 10$) provides a better match to the observations than an unclumped model,  The overall line intensity distributions  are consistent with a freely expanding, spherically symmetric  1/$r^2$ outflow with constant ionization fraction and modestly clumped density structure. 

\end{abstract}


\section{Introduction}

$\gamma$~Velorum is a binary system comprising a WC8-type Wolf-Rayet \citep{Smith68b} and O7.5~III companion \citep{DeMarco99}. The binary is well-defined, with a period of 78.53 days, $i \sim 65^{\circ}$ \citep{Schmutz97} and a separation of $\sim$3.8~milliarcsec (\citealt{North07}; \citealt{Millour07}). The O-star appears typical and is the more massive star in the binary with $M_{\mathrm{O}}$~=~28.5$M_{\odot}$ against $M_{\mathrm{WR}}$~=~9.0$M_{\odot}$ (\citealt{Conti72}; \citealt{DeMarco99}; \citealt{North07}). The WC8 is the nearest example of a Wolf-Rayet star, at a distance of 342~pc (see below; \citealt{vanLeeuwen07}), affording  the greatest opportunity to study this important class in depth

The outflow from the WC8 is considerable, with a mass-loss rate of $2.95{\times}10^{-5} \leq {M_\odot}$yr$^{-1} \leq 8.85{\times}10^{-5}$ if unclumped (and a factor of 3.8 lower if clumped by a factor of 10), at least 100 times that of the O-star ( \citealt{DeMarco00}; Barlow, Roche \& Aitken 1988). The ejecta can be detected at great separations from the star, several times the binary separation, from the Bremsstrahlung continuum at radio wavelengths \citep{Seaquist76} and from low-density forbidden emission lines (e.g. \citealt{Barlow88}). The wind terminal velocity, found from those forbidden emission lines, is $\sim$1520~kms$^{-1}$, confirmed by UV eclipse absorption line profiles (St-Louis, Willis \& Stevens 1993). An energetic wind-collision zone is detected in UV and X-rays, with variability consistent with the effect of the alignment of the binary to the observer (\citealt{StLouis93}; \citealt{Stevens96}; \citealt{Schild04}). X-ray absorption column densities suggest a lower mass-loss rate by an order of magnitude, which may be evidence of clumping in the inner wind \citep{Schild04}.

Mid-infrared spectroscopy of $\gamma$~Vel has been used to investigate the wind structure.  One important motivation was to determine the neon abundance in the wind, as Wolf-Rayet stars are thought to be the origin of most $^{22}$Ne in the ISM. This was done from the forbidden line emission [{Ne\,\sc ii}] at 12.81-$\mu$m and [{Ne\,\sc iii}] at 15.56-$\mu$m (\citealt{Barlow88}, van der Hucht et al  1996) both of which have wide, flat spectral profiles suggesting that much of the emission arises from the outer regions where terminal velocity has been reached.  However, from an uncertainty in the mass-loss rate due to the uncertain distance, the Ne enrichment was not initially found \citep{Barlow88}. Further analysis in the mid-IR and the introduction of wind-clumping models suggests that Ne is enriched in WC star winds (\citealt{Dessart00}; \citealt{Smith05}; \citealt{Ignace07}). 

In order to correctly calibrate the spatial scale of the observed outflow it is necessary to fix the distance to the object. The distance to $\gamma$~Vel has been the subject of many revisions in recent years. In this paper we use the revised Hipparcos parallax of 2.92$\pm$0.3~mas from \citet{vanLeeuwen07}. This translates to a distance of 342$^{+39.2}_{-31.9}$~pc, which is broadly consistent with the interferometry-derived distances of 336~pc and 368~pc estimated by \citet{North07}  and \citet{Millour07} respectively.  This distance is substantially greater than the initial Hipparcos estimate of 258~pc. 

In this paper, we present spatially resolved 8-13~$\mu$m spectroscopy and narrow-band mid-infrared images of the $\gamma$~Vel outflow and investigate the density structure of the wind. 

\section{Data}

\begin{table}
\centering

\begin{tabular}{@{}llccc@{}}
\hline
 Date & Filter & $\lambda_{centre}$ & $\lambda_{50\%}$ & On-source \\ 
(UT) & name & ($\mu$m) & ($\mu$m) & time (min) \\
\hline\hline
\multicolumn{4}{l}{\itshape Imaging} \\
2006 Apr 10 & [{Ne\,\sc ii}] & 12.81 &  12.68--12.91 & 34.5 \\
2006 Apr 18 & [{Ne\,\sc ii}]cont & 13.10 & 12.95--13.17 & 34.5 \\
2006 Apr 27 & [{S\,\sc iv}] & 10.52 & 10.37--10.54 & 33.4 \\
2006 May 06 & [{Ar\,\sc iii}] & ~~8.99 & 8.93--9.06  & 33.4 \\
\hline
\multicolumn{4}{l}{ \itshape Spectroscopy, slit PA=0$^{\circ}$} \\
2006 May 22 & N broad & 10.36 & & 16.3 \\
 \hline
\end{tabular}
\caption{Log of T-ReCS imaging and spectroscopy of $\gamma$~Vel}
\label{tab:gamveldata}
\end{table}

The imaging and spectroscopic observations of $\gamma$~Velorum were obtained with the Gemini Thermal-Region Camera Spectrograph (T-ReCS) \citep{Telesco98} in April and May 2006 under Gemini programme GS-2006A-Q-64. The instrument uses the telescope in the chop-nod mode, with a maximum chop-throw of 15$^{\prime\prime}$, used for all of the observations presented here. The fixed pixel size of T-ReCS is 0.09$^{\prime\prime}$. Images of the WR-O binary and a standard star (SAO~219422) were taken in four filters covering the [{Ne\,\sc ii}]  and  [{S\,\sc iv}] emission lines and nearby continuum regions (see Table \ref{tab:gamveldata}).    Spectra were obtained through a 0.35$^{\prime\prime}$ slit and the broad-$N$ filter, centred at 10.36-$\mu$m and with 50\% transmission between 7.70-12.97~$\mu$m.  The spectral resolution was 0.08$\mu$m, sampled at  0.022 $\mu$m/pixel. Three exposures were taken on the night, with the total on-source time given in Table \ref{tab:gamveldata}. Observing conditions were good apart from the night of the [{Ar\,\sc iii}] imaging observations, which caused some striping in the final image, and the spectra were obtained at rather high airmass as the target set.  A log of the  imaging and spectroscopic observations is provided in Table \ref{tab:gamveldata}.

Data reduction was completed using the T-ReCS {\sc iraf} pipeline. The chop frames were mean-averaged and the alternating positions subtracted to remove residual background structure.  Examination of the spatial-axis distribution of the flux suggested there would be no significant off-source emission within the chop throw. The spectra were centred and aligned along along one pixel-row, though  the corrections applied were minimal. A wavelength calibration was applied using the sky emission line list supplied by Gemini for the $N$-band. The resulting spectra were then averaged together to increase the signal-to-noise ratio of the $\gamma$ Vel spectrum. Similar processing was applied to the standard star spectrum of HD89998. Telluric-correction and flux-calibration were performed by dividing the averaged $\gamma$ Vel by the standard star spectrum and multiplying by a flux-calibrated blackbody to match the temperature of the star; 4280~K derived from the spectral-type of K1{\sc iii} for HD89998 (SIMBAD; \citealt{Johnson66}). A 1-D extraction of the spectrum was taken, summing the spatial-axis rows, and, separately, the individual rows along the spatial-axis of the 2-D spectrum were binned into 3-pixel bins, mean-averaging the rows to produce the binned spatial spectra.

\section{Results}

\subsection{Spectroscopy}

\begin{figure}
\begin{center}
\includegraphics[width=11cm]{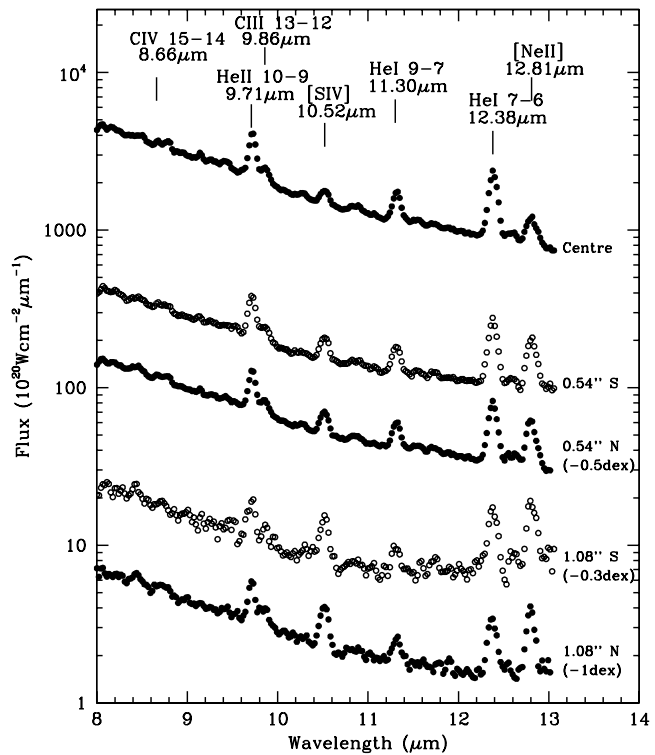}
\caption{Spectra of $\gamma$~Velorum at the peak mid-IR flux and at spatial separations to the North and South. The [{S\,\sc iv}] and [{Ne\,\sc ii}] emission line components at 10.5~ and 12.8~$\mu$m increase in prominence away from the central position.} 
\label{fig:gamvelspec}
\end{center}
\end{figure}

\begin{table}
\centering
\begin{tabular}{@{}lll@{}}
\hline
~$\lambda_{obs}$ & Identification & Flux \\
($\mu$m) & & (10$^{-20}$Wcm$^{-2}$) \\
\hline\hline
8.25 & C{\sc iv}~19-17, C{\sc III}~28-20 & $<$10 \\
8.48 & C{\sc iii}~16-14, C{\sc iii}~21-17 & $<$15 \\
8.67 & He{\sc i}~14-8, C{\sc iii}~25-19, C{\sc iv}~15-14 & $<$20 \\
8.78 & He{\sc i}~10-7, He{\sc ii}~20-14 & $<$40 \\
9.72 & He{\sc ii}~10-9, He{\sc ii}~13-11, C{\sc iv}~20-18 & 179.8 \\
9.86 & C{\sc iii}~13-12, He{\sc i}~18-9, He{\sc ii}~25-16 & 63.2 \\
10.52 & [{S\,\sc iv}], He{\sc i}~12-8, He{\sc ii}~21-15 & 54.4 \\
& He{\sc ii}~24-16, C{\sc iii}~20-17, C{\sc iv}~16-15 & \\
11.32 & He{\sc i}~9-7, He{\sc ii}~23-16, He{\sc ii}~18-14 & 66.4 \\
& He{\sc ii}~16-13, C{\sc iv}~21-19 & \\
12.38 & He{\sc i}~7-6, He{\sc i}~11-8, He{\sc ii}~14-12 & 155.6 \\
 & He{\sc ii}~22-16 & \\
12.58 & He{\sc i}~14-9 & 13.4 \\
12.81 & [Ne\,\sc ii], C{\sc iv}~17-16 & 53.4 \\
\hline
\end{tabular}
\caption{Wavelengths of prominent emission lines together with the most important components  in the blended lines. The $\lambda$ and fluxes are from Gaussian fits to lines in the centre bin. Errors on the fluxes are estimated to be at most 10\% of the flux of the most prominent lines.}
\label{tab:gamvelemlines}
\end{table} 

Cuts across the spatial-axis of the spectrum of $\gamma$~Vel yield  FWHM of 0.45$^{\prime\prime}$ in the continuum emission between 10 and 12~$\mu$m, which matches the spectroscopic standard star PSF and indicates an unresolved core emission source. There are several prominent emission line blends present in the spectrum. These are listed in Table \ref{tab:gamvelemlines}, together with a number of weaker emission lines and the principal lines contributing to each blend. Five of the most prominent lines are attributed to helium or carbon recombination line blends, but the lines at 10.52 and 12.81~$\mu$m have contributions from  the ground state fine-structure  S$^{3+}$ and Ne$^{+}$ transitions respectively,  blended with helium and/or carbon, see Table \ref{tab:gamvelemlines}. In the WC8 star, hydrogen lines are not expected and are not seen in optical or UV spectra (e.g. Willis 1982).  Ionized carbon is, of course, an important emission component, with  {C\,\sc iii} and {C\,\sc iv} recombination lines previously observed in the near- and mid- IR (e.g. \citealt{Aitken82}) and UV (\citealt{Castor72}; \citealt{Hucht73}). The blended line fluxes measured in the peak bin are listed in Table \ref{tab:gamvelemlines} and the spatially-binned blended line fluxes are provided in Table \ref{tab:alllinefluxes}.

\begin{figure}
\includegraphics[width=8.8cm]{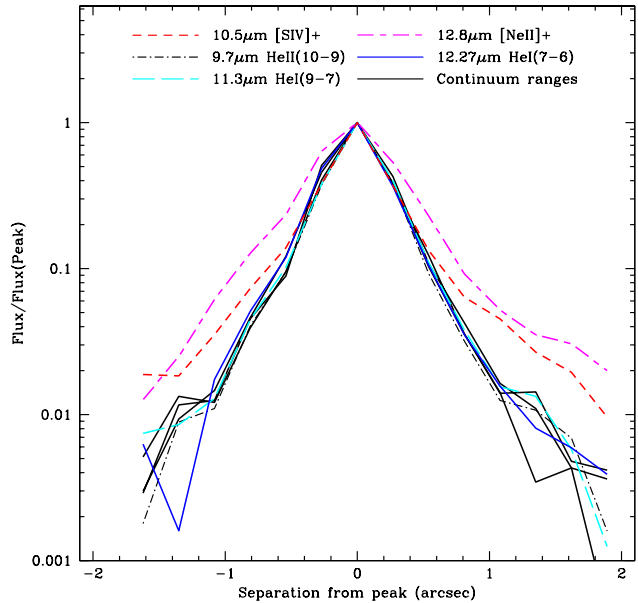}
\caption{Normalised spatial profiles of emission lines and continuum from $\gamma$~Vel. The continuum emission at a number of wavelengths  between 10 and 12~$\mu$m is shown by the thin solid lines and has a very similar profile to the those of the recombination lines, illustrated by the  He{\sc i} 7-6 and 9-7 lines and the He{\sc ii} 10-9 line.  These are unresolved,  but the lines at 10.5 and 12.8$\mu$m, which contain contributions from [S {\sc iv}] and [Ne {\sc ii}] have more extended profiles.    Negative separations are to the South, positive to the North. } 
\label{fig:lineflux}
\end{figure}

\begin{figure}
\includegraphics[width=8.8cm]{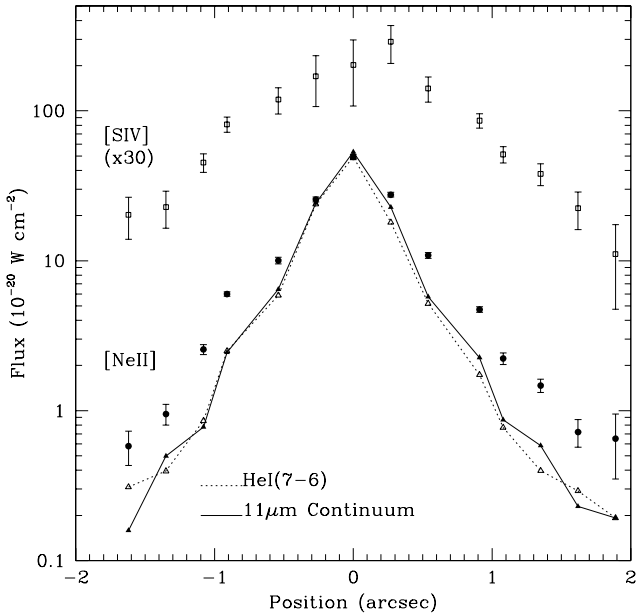}
\caption{Spatially-resolved, extended emission from the [{S\,\sc iv}] and [{Ne\,\sc ii}]  forbidden lines in the $\gamma$~Velorum outflow obtained from the spectra.  Unresolved emission from the inner wind is  illustrated by the 11$\mu$m continuum and the 12.37$\mu$m emission line, the latter attributed primarily to the HeI (7-6) line (closed and open triangles respectively) and normalised to the [{Ne\,\sc ii}] emission peak.  The extended  [{S\,\sc iv}] and [{Ne\,\sc ii}] emission is shown after the line deblending described in the text.} 
\label{fig:forblineflux}
\end{figure}

The fact that the recombination line blends exhibit comparable spatial profiles to the continuum emission, with little dependence on wavelength, indicates that seeing contributes substantially to the profiles and masks the effects of diffraction across the 9.7 to 12.4 $\mu$m wavelength range; over this range, the diffraction-limited image width increases from 0.26 to 0.34 arcsec.  The continuum and recombination line emission is centrally concentrated from an unresolved emission source, arising  from the dense, inner regions of the wind. The lines at 10.5 and 12.81-$\mu$m show a different, spatially-extended emission profile from the stellar continuum and He lines. The blended and deblended (see below) emission is plotted in Fig. \ref{fig:lineflux} and Fig. \ref{fig:forblineflux} respectively. 

The [{S\,\sc iv}] emission is blended with helium and carbon emission at 10.5-$\mu$m and the [{Ne\,\sc ii}] emission is blended with {C\,\sc iv}~17-16 at 12.77$\mu$m (Table \ref{tab:gamvelemlines}). The 10.5-$\mu$m blend shows a FWHM comparable to the unresolved continuum, reflecting the dominance of the He/C contribution to that line in the inner region. However, beyond $\sim$0.5$^{\prime\prime}$ the line flux decreases more slowly than the continuum, indicating resolved emission from [{S\,\sc iv}] at large radii in the wind. The 12.81-$\mu$m emission is less heavily blended. The largest contribution is from [Ne {\sc ii}]  while the contribution from {C\,\sc iv} emission is weak, as can be seen from the small flux in the 8.66-$\mu$m line - the strongest {C\,\sc iv} emission line in this spectral range. The 12.8$\mu$m line blend has a  FWHM =0.65$^{\prime\prime}$, significantly larger than those of the continua and 10.5~$\mu$m emission line blend, which have FWHM in the range 0.43 to 0.48 $^{\prime\prime}$.  This reflects the dominant contribution of  [{Ne\,\sc ii}] in the 12.8~$\mu$m blend and indicates an extended distribution. Careful correction for He and C lines at 10.5$\mu$m is required to extract the [S {\sc iv}] distribution. 

\begin{figure}
\begin{center}
\includegraphics[width=8.8cm,bb=10 80 600 660]{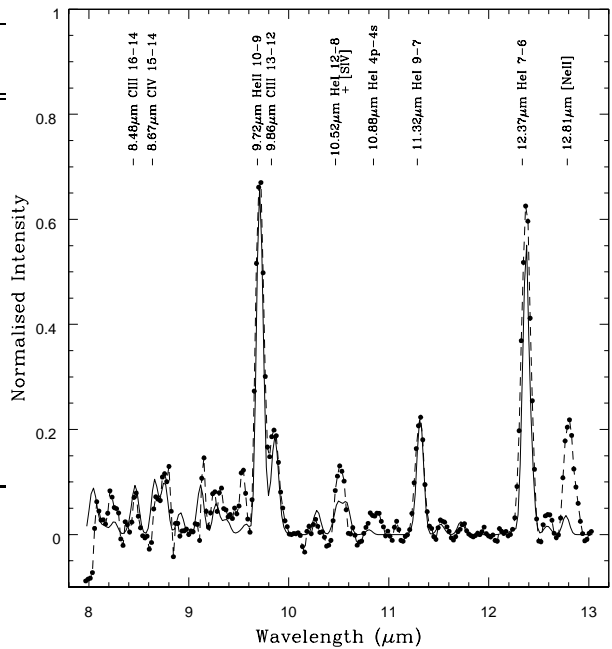}
\caption{The continuum-subtracted spectrum of  $\gamma$~Vel extracted at 0.27" S of the peak (circles connected by the dashed line) overlaid by the scaled synthetic spectrum (solid line).  The most prominent emission line blends are labelled with the wavelength and the dominant contributor, but several more components are also important in many cases - see Table 2.}
\label{fig:synthspec}
\end{center}
\end{figure}

To deblend the lines, the {\sc intrat} program (supplied by Prof P. Storey) was used to compute line ratios between the different helium and carbon transitions, allowing for those components of the 10.5 and 12.81-$\mu$m emission blends to be calculated and subtracted. The reference lines against which ratios of each species would be computed, were {He\,\sc i}~7-6, {He\,\sc ii}~10-9, {C\,\sc iii}~13-12 and {C\,\sc iv}~15-14, each the strongest line of their species in the 8-13~$\mu$m spectral region. {\sc intrat} calculates the line flux ratios for hydrogenic species in specified densities and temperatures using case B recombination \citep{Baker38}, which is justified for the opaque inner wind where the lines are thought to form (e.g. \citealt{Stevens96}).  The electron temperature and density adopted were 30,000~K and 10$^{10}$ cm$^{-3}$ respectively, but the dependence of the line ratios on these quantities is fairly weak.   
The contributions from weak lines were estimated by generating a synthetic spectrum from the line ratios generated from {\sc intrat} and fitting to the peak flux spectrum,  where the He and C emission lines  are strongest. The contributions from {He\,\sc i}, {He\,\sc ii}, {C\,\sc iii} and {C\,\sc iv} were adjusted to give a good match to the most prominent emission lines.   Once the ratios of {He\,\sc i}, {He\,\sc ii}, {C\,\sc iii} and {C\,\sc iv} to each other were determined, they were applied to the measured fluxes at offset positions  (see Fig. \ref{fig:synthspec}). The synthetic spectrum was broadly consistent with the offset spectra, giving confidence that  the He and C line contributions to the 10.52 and 12.81-$\mu$m lines could be estimated.  

\begin{table}
\centering
\begin{tabular}{llllll}
\hline
Position & \multicolumn{5}{c}{Flux (10$^{-20}$Wcm$^{-2}$)}  \\
(arcsec) & 9.72~$\mu$m  & 10.52~$\mu$m & 11.30~$\mu$m & 12.37~$\mu$m & 12.81~$\mu$m\\
\hline\hline
-1.62 & 0.322 & 1.023 & 0.493 & 0.977 & 0.680 \\
-1.35 & 1.598 & 1.003 & 0.567 & 0.249 & 1.347 \\
-1.08 & 1.975 & 1.919 & 0.854 & 2.705 & 3.318 \\
-0.81 & 7.263 & 3.974 & 2.962 & 7.914 & 6.836 \\
-0.54 & 17.05 & 7.555 & 6.763 & 18.54 & 12.61 \\
-0.27 & 67.15 & 20.68 & 25.14 & 75.38 & 33.84 \\
0 & 179.8 & 54.43 & 66.41 & 155.6 & 53.42 \\
0.27 & 69.02 & 20.15 & 26.44 & 56.92 & 28.43 \\
0.54 & 16.61 & 7.263 & 7.305 & 16.38 & 12.20 \\
0.81 & 5.738 & 3.472 & 2.526 & 5.480 & 4.895 \\
1.08 & 2.241 & 2.464 & 1.031 & 2.440 & 2.769 \\
1.35 & 1.931 & 1.456 & 0.887 & 1.258 & 1.871 \\
1.62 & 1.240 & 1.060 & 0.384 & 0.920 & 1.627 \\
1.89 & 0.286 & 0.524 & 0.0829 & 0.607 & 1.064 \\
\hline
\end{tabular}
\caption{Spatially-measured line fluxes of the prominent emission line blends through the 0.35 arcsec wide T-ReCS slit. The detector pixels were binned in threes in the spatial direction, giving spaxels of 0.27 arcsec. Negative separations are to the South, positive to the North.}
\label{tab:alllinefluxes}
\end{table}

Inspection of Fig. \ref{fig:synthspec} shows that the spectral structure between 10 and 12~$\mu$m, which results from high-{\it n} recombination transitions  is well matched by the synthetic spectrum. The region between 8 and 10~$\mu$m is noisier because of the poorer atmospheric transmission.  Note that the dominant contribution to the line at 10.88~$\mu$m is attributed to the 4p-4s transition in HeI, which is not a hydrogenic transition.  The  {C\,\sc iii} and {C\,\sc iv} lines  are relatively weak, resulting in significant uncertainties  which propagate into the contributions from the higher {\it n} carbon lines at 10.5~$\mu$m and 12.8$\mu$m.  About 2/3 of the 12.37~$\mu$m and 9.72~$\mu$m line fluxes are attributable to the He I (7-6) and HeII (10-9) components respectively.  [{S\,\sc iv}] emission was found to contribute only $\sim$15\%  of the total 10.5-$\mu$m emission at the peak position, but increases to $\sim$50\% at 0.54$^{\prime\prime}$ from the centre. The recombination line contributions  were subtracted from the spatially-separated 10.5-$\mu$m flux measurements by subtracting the scaled synthetic spectrum.  The resulting spatial profiles are plotted in  Fig. \ref{fig:forblineflux}.   

Because the contribution from [{S\,\sc iv}] to the 10.52~$\mu$m emission blend is small, the peak [{S\,\sc iv}] flux  is very uncertain,   This makes the FWHM of the [{S\,\sc iv}] emission difficult to estimate, but it is nonetheless clearly extended with a much flatter distribution than the continuum emission.   [{Ne\,\sc ii}] emission was found to contribute $\sim$88$\pm 5\%$ of the 12.81-$\mu$m flux at the peak, increasing to 95\% at 0.54$^{\prime\prime}$. The emission is spatially centrally peaked and symmetrical within the estimated errors, with a FWHM$\sim$0.6$^{\prime\prime}$ and a steeper profile than [{S\,\sc iv}].

\subsection{Imaging}

After processing the imaging data,  images in the four filters, [{S\,\sc iv}], [{Ar\,\sc iii}], [{Ne\,\sc ii}] and [{Ne\,\sc ii}]-cont, were produced of  $\gamma$~Vel and a nearby star, SAO  219422, which provided an estimate of the PSF.  The images were obtained on a number of different nights and in a variety of conditions.  The standard star images generally showed some low level departures from circular symmetry, presumably due to optical aberrations,  and  similar structures were visible in the $\gamma$~Vel imaging.  The PSFs derived from the standard star images have FWHM $\sim 0.4$ to 0.6 arcsec.  The PSF in both the standard star and  $\gamma$~Vel  images is typically more centrally concentrated than that of the continuum measured from the spectra, which were obtained at higher airmass and in relatively poor seeing conditions;  only the images of the WR and standard in the  [{Ar\,\sc iii}] filter show a comparable spatial PSF to the WR spectra.

To obtain images of the spatial distribution of the [{S\,\sc iv}] and [{Ne\,\sc ii}] emission lines, the underlying continuum and recombination line emission from the WR star had to be subtracted. This continuum was to have been obtained from the [{Ar\,\sc iii}] and [{Ne\,\sc ii}]-continuum filter images, which sample the continuum emission near to the wavelengths of the forbidden-line emission. The [{Ar\,\sc iii}] filter does not include any prominent emission lines but the   [{Ne\,\sc ii}]-continuum filter includes a contribution from the {He\,\sc ii}~11-10 transition at 13.12-$\mu$m. The contribution from this line to the flux contained within the [N{e\,\sc ii}]-continuum filter was estimated to be $\sim$6\% using the measurements in \citet{Aitken82} and the transmission function of the filter. 

Unfortunately, the continuum subtractions did not appear satisfactory, despite using the star spectral energy distribution to normalize the continuum fluxes to the wavelength of the forbidden emission lines. It seems that the variable PSF between the observations dominates and leaves substantial residuals, although other factors (e.g. filter leaks or ghosts) may also contribute. The spectra indicate that while the contribution from the [{Ne\,\sc ii}] emission line to the total flux measured within the [{Ne\,\sc ii}] filter is about 15\% in the central 0.3 arcsec of the image, the contribution from the [{S\,\sc iv}] line is only about 3\%. The approach used in the end was to smooth the images and subtract the continuum images with the aim of scaling the radial flux to best match the spatial flux distribution measured in the spectra.  The [{S\,\sc iv}] subtraction was more troublesome, so the [{S\,\sc iv}] image of the standard was scaled to the spectrum flux spatial distribution and subtracted, but even then the uncertainties are considerable. The inner part of the profile is affected by the scaling of the continuum image and the outer regions by the adopted  background level. 

\begin{figure*}

\includegraphics[width=6.6cm]{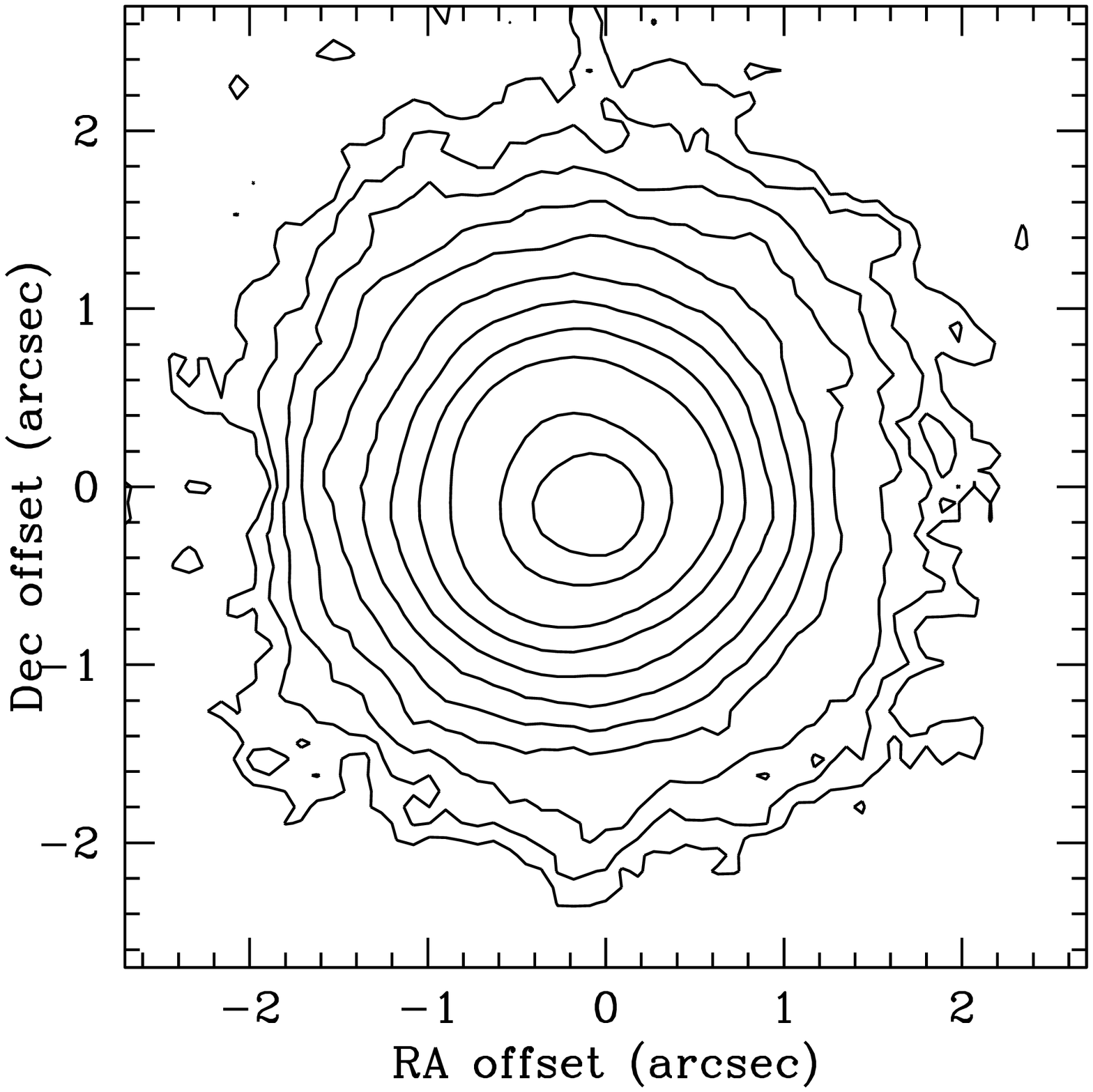}
\includegraphics[width=6.6cm]{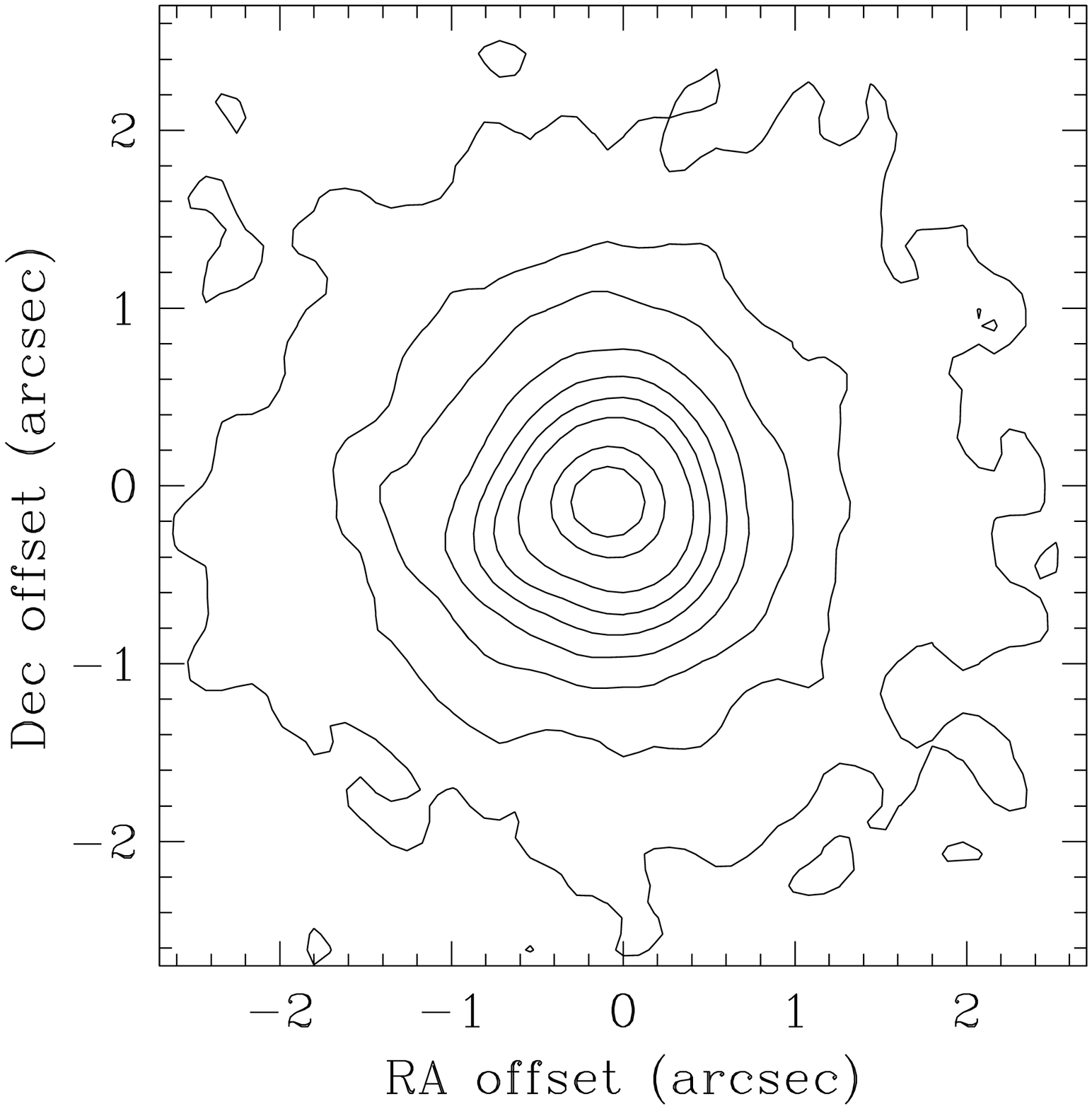}
\includegraphics[width=10.7cm, bb= 100 200 550 560, angle=-90]{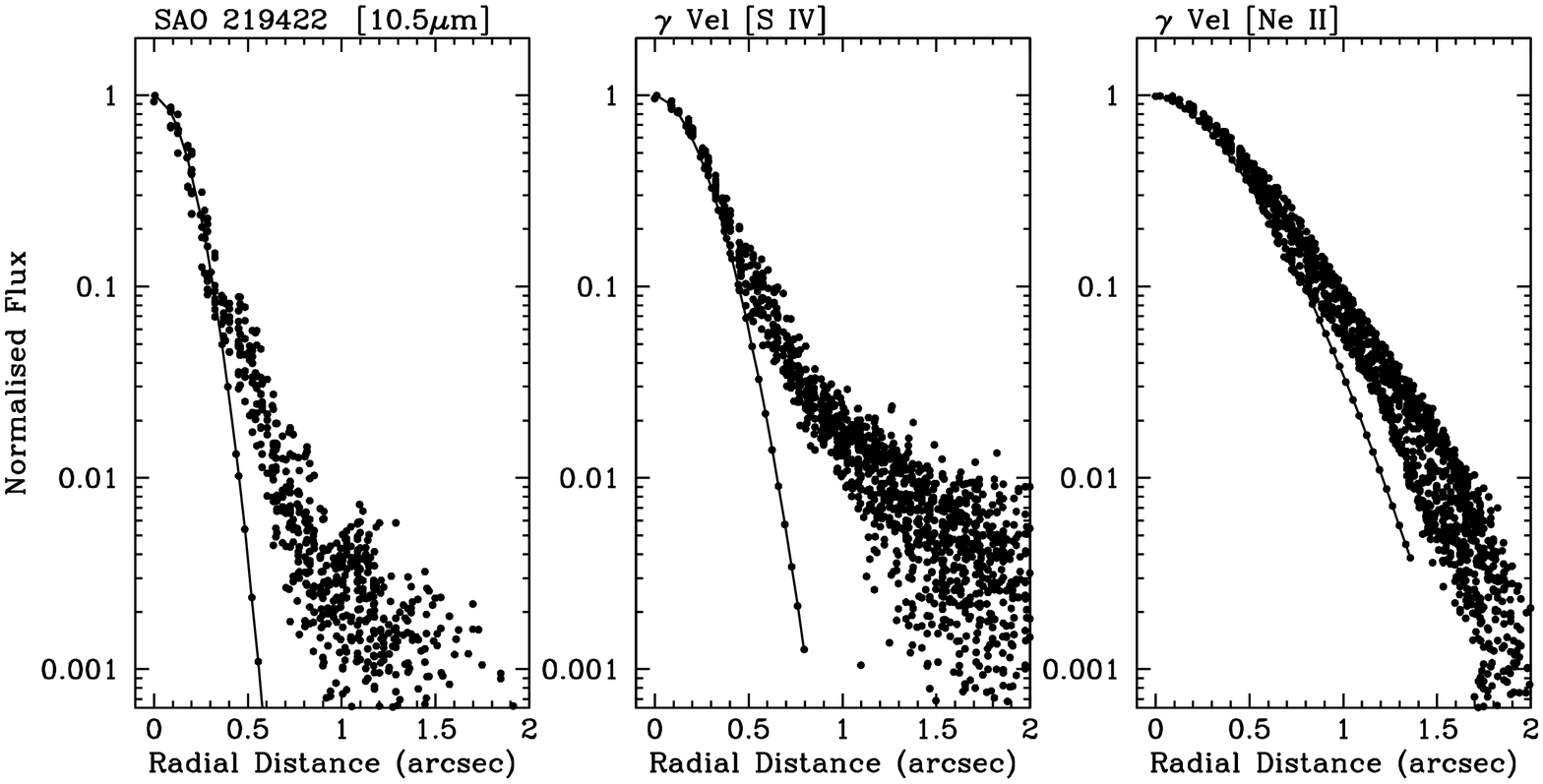}
\caption{Top: Contour plots of the [{Ne\,\sc ii}] (left) and [{S\,\sc iv}] (right) line intensities. North is up, East to the left.  The contours are  logarithmic in 0.2dex intervals. Bottom: Radial profiles of the comparison star, SAO 219422 at 10.5$\mu$m and the  [{S\,\sc iv}] and [{Ne\,\sc ii}] line intensities in  $\gamma$ Vel.  The solid  lines are gaussian fits to the core emission. Because of the uncertainties in subtraction of recombination line and continuum emission, the flux levels and distribution in the cores of the line images and profiles may be subject to systematic errors}

\label{fig:gamvelcontour}
\end{figure*}

The continuum-subtracted  [{Ne\,\sc ii}] image,    Fig. \ref{fig:gamvelcontour}, is centrally peaked with emission detected to $\sim$ 2 arcsec from the centre. The contours  are largely  centro-symmetric, but with slightly enhanced emission to the East. The FWHM of the [{Ne\,\sc ii}]  image  $\sim$0.9$^{\prime\prime}$, but this figure is rather uncertain because of the potential errors in continuum subtraction, and is greater than the 0.65 arcsec FWHM obtained from the spectra.  A contour plot of the [{S\,\sc iv}] emission is also shown in Fig. \ref{fig:gamvelcontour}. Extended low-level emission is apparent, again with indications of enhanced emission to the East.  Note however that there appears to be residual emission from the compact core in the [{S\,\sc iv}]  image that is not present in the spatial distribution of the [{S\,\sc iv}]  emission derived from the spectra. This leads to a compact FWHM comparable to that of the PSF standard star at $\sim$0.45 arcsec.  Radial profiles generated from the  [{S\,\sc iv}] and [{Ne\,\sc ii}]  images are displayed together with the profile of the comparison star SAO 219422 in the 10.5$\mu$m filter.  Extended emission in the forbidden lines is clearly present beyond the core, with the  [{S\,\sc iv}] profile falling more slowly than  [{Ne\,\sc ii}] beyond 0.5 arcsec. 

Whilst the detailed forms of the radial profiles are subject to substantial uncertainties, all the trial images produced retain a high degree of azimuthal symmetry indicating that a spherically symmetric outflow is a good approximation.  

\section{Wind model}

The spatially extended emission from the fine structure lines provides a potential diagnostic of the wind structure. The ratio of the [{Ne\,\sc ii}] to [{S\,\sc iv}] line intensity probes the wind region with densities near the critical densities for the transitions, while the radial profiles of these lines will trace the level populations, ionic abundances  and density of the wind. 

A simple wind model was constructed to compare the observed peak and extended emission from [{S\,\sc iv}] and [{Ne\,\sc ii}] in the wind with expectations of a freely expanding envelope with constant ionization fraction. Two model winds were used: (i) a smooth wind as described by \citet{Barlow88}, but modified by the revised distance of 342~pc  and a corresponding mass loss rate of $\dot{M}$ = $5.7 {\times}10^{-5} {M_\odot}$  yr$^{-1} $ (rather than 460~pc and the mass loss rate of $\dot{M}$ = $8.8 {\times}10^{-5} {M_\odot}$  yr$^{-1} $ adopted in \citet{Barlow88}); and (ii) a clumped wind with a clumping factor $f= 10$ and a mass loss rate of $\dot{M}$ = $1.5 {\times}10^{-5} {M_\odot}$  yr$^{-1} $, corresponding to the values determined by de Marco et al (2000) adjusted for the distance adopted here of 342~pc.  The models  calculate the flux  admitted  by the T-ReCS slit from a spherical 1/$r^{2}$ distribution of gas with concentric shells of decreasing electron density, $n_{e}$, away from the core  for an optically-thin (at the wavelengths of line emission) wind.  The instrument slit was projected onto the wind model and intervals along the spatial-axis of the slit were defined to create bins, representing the pixel scale of the observing instrument. The volume of each of the individual density shells visible to each bin was then calculated. Finally the mean fractional upper atomic level population times the radius interval ($\bar{f}_{u}{\delta}r$) for each density sphere, shown to trace the observed line flux, from Table 4 in \citet{Barlow88} was multiplied by the normalized volume of the corresponding sphere observed in each bin to reproduce the flux observed in each bin.

The collision strengths for  [{S\,\sc iv}] and [{Ne\,\sc ii}] adopted are 8.27 and 0.30 respectively (Saraph \& Storey 1999, Johnson \& Kingston 1987). The  Einstein A Values of $7.74  {\times}10^{-3}$ and $8.85  {\times}10^{-3}$ then indicate critical densities n$_{crit}$ {of $3.4 {\times}10^{4}$ cm$^{-3}$ for [{S\,\sc iv}]  and an order of magnitude greater at $5 {\times}10^{5}$ cm$^{-3}$ for [{Ne\,\sc ii}].  As discussed by \citet{Barlow88}, line formation peaks in the region in the wind where n$_{e} \sim n_{crit}$.    For the two cases considered here, n$_{crit}$ occurs at radii of $2.7 {\times}10^{15}$ cm and $7.1 {\times}10^{14}$ cm for  [{S\,\sc iv}] and [{Ne\,\sc ii}] for model (i) and  at radii of $4.4 {\times}10^{15}$ cm and $1.1 {\times}10^{15}$ cm for  [{S\,\sc iv}] and [{Ne\,\sc ii}] for model (ii). 

\begin{figure}
\begin{center}
\includegraphics[width=9cm]{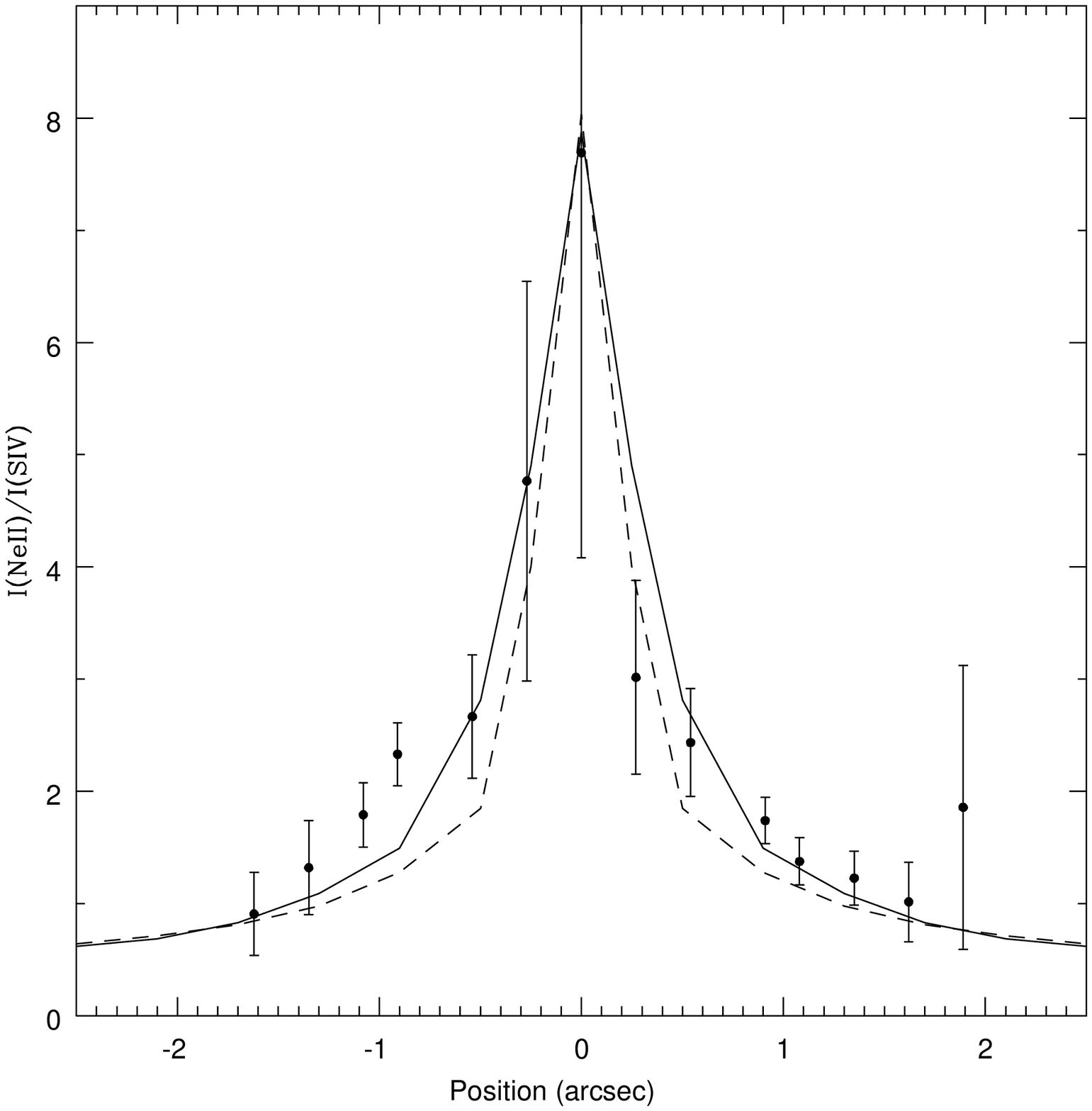}
\caption{The  [{Ne\,\sc ii}] to [{S\,\sc iv}]  line ratio (data points) and the line ratios calculated from the  wind models.   The dashed line represents model (i)., the unclumped wind, and the solid line represents model (ii) the clumped wind with $f= 10$. }
\label{fig:linefluxratio}
\end{center}
\end{figure}

Because of its lower critical density,  the [{S\,\sc iv}] emission is expected to peak further out in the wind than the [{Ne\,\sc ii}] emission, and for a constant ionization wind the ratio of the line fluxes will vary until the density intersected by the instrument beam falls below the critical density of [{S\,\sc iv}], beyond which  it should approach a constant value given by the atomic parameters and the relative ionic abundances of the two species. 

The observed  [{S\,\sc iv}] and [{Ne\,\sc ii}] emission line distributions given by the models  can be compared with the spatial distributions of the emission lines extracted from the spectra.   Because they were observed simultaneously, the  emission line  and continuum distributions extracted from the spectrum are more reliable than those obtained from the images, which were taken at different times under a range of conditions.

The observed spatial dependence of the  [{Ne\,\sc ii}] to [{S\,\sc iv}] line intensity ratio is shown together  with the ratio  calculated from the two models in Figure  \ref{fig:linefluxratio}.  There is reasonable qualitative agreement with the observations, although the models cannot account for the observed asymmetry and neither model matches the observations in detail.  In model (ii), the clumping factor of 10 increases the emitting density of the wind so that the regions of critical density occur at greater distances from the centre, even though the mass loss rate is substantially lower. This means that the [{Ne\,\sc ii}] to [{S\,\sc iv}] line ratio approaches the asymptotic value beyond 2 arcsec, whereas in the unclumped wind model, the critical density is reached at a radius smaller by a factor of 1.6 and so the line ratio drops more steeply in the inner region and starts to flatten out beyond 1 arcsec.    Furthermore, because the peak intensity of the [{S\,\sc iv}] line is very uncertain, the line ratio at the centre has a correspondingly large error, which makes the normalization of the fits very uncertain. 

 In the outer wind,  where the density falls below the critical densities of the transitions, the  [{Ne\,\sc ii}] to [{S\,\sc iv}] line intensity ratio approaches 1.0 $\pm$0.2.   The dominant ionisation stage of neon is  Ne$^{2+}$,  so the observations presented here cannot place useful constraints on the overall neon abundance.    However, they can provide estimates of  the ${\rm Ne}^{+}$ to ${\rm S}^{3+}$  ionic ratio. The emission line intensity  I$_\lambda \propto {\rm h}\nu \int{n_{u}~{\rm A}_{2,1}~{\rm V}}$, and as the Ne and S lines arise from the same volume, the ratio of the line intensities provides an estimate of the  ${\rm Ne}^{+} /{\rm S}^{3+}$ abundance ratio.  The fractional upper level populations from Table 3 of  \citet{Barlow88} and the older atomic datasets used in that work  would have given an ionic abundance  ratio ${\rm Ne}^{+} /{\rm S}^{3+} \sim$ 50, while the more recent atomic datasets for  ${\rm Ne}^{+}$ and  ${ \rm S}^{3+}$ from the CHIANTI database (www.chianti.rl.ac.uk) yield a  ${\rm Ne}^{+} /{\rm S}^{3+}$ ratio of 67 in the outer wind.

Fig. \ref{fig:linefluxmodel} compares the spatial distribution of the [{Ne\,\sc ii}] and  [{S\,\sc iv}] emission lines with the model predictions.  The spherical wind and density-dependent emission of the forbidden lines does reasonably reproduce the observed emission from a separation of 0.6$^{\prime\prime}$ outwards, although the range of scales sampled is small.  The slope of the emission past $\sim$0.6$^{\prime\prime}$ appears to confirm that a spherical wind geometry with the 1/$r^2$ density-dependent emission distribution is a good match for the observed data in the outer wind.  However, the  models diverge in the inner regions. In the unclumped wind model,  the line intensities continue to increase steeply towards the central pixel as the critical densities are reached relatively close to the centre.  The  [{S\,\sc iv}] emission from the clumped wind model starts to flatten at a radius of $\sim1$arcsec, providing a better match to the observations;  the model indicates that the [{Ne\,\sc ii}]  emission line profile would flatten inside a radius of 0.3 arcsec, but unfortunately this regime is not well-sampled by the data.     

\begin{figure}
\begin{center}
\includegraphics[width=10cm, bb=70 200 600 560]{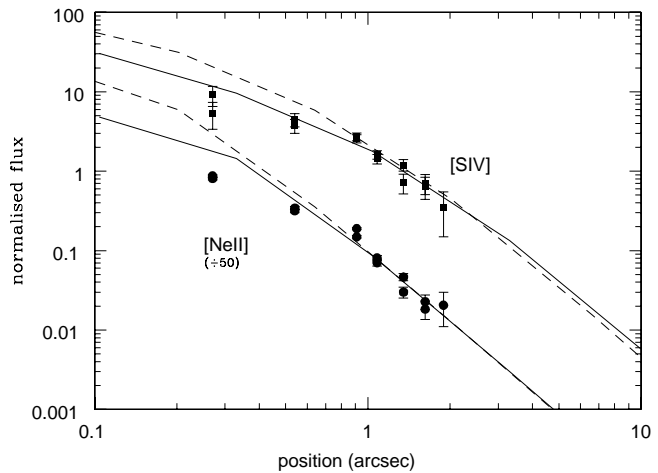}
\caption{Emission observed in the [{S\,\sc iv}] and [{Ne\,\sc ii}] lines (data points) and the emission distributions calculated from the  wind models. dashed lines represent the unclumped and solid lines the clumped wind profiles.}
\label{fig:linefluxmodel}
\end{center}
\end{figure}

The clumped wind provides a better overall match to the spatial dependences of the emission line  intensity distributions and the [{Ne\,\sc ii}] to [{S\,\sc iv}] line ratio.  A modest clumping factor  $f \sim 10$ provides a reasonable match to the data; a more highly clumped wind might improve the fit somewhat, but a very highly clumped wind   (with $f  >>$10) would  flatten the  profiles further and start to diverge from the observations.

The models made a number of assumptions that could have an effect.   The emission was assumed to be optically-thin at all radii. As discussed by \citet{Barlow88}, the fine-structure lines are expected to be optically thin throughout the outflow, but the infrared free-free continuum is expected to become  optically thick in the inner dense regions.  The radius at which the continuum becomes optically thick can be estimated from the work of Wright and Barlow (1975) who provide an expression for the characteristic radius of the emitting region in the outflow. For $\gamma$~Vel, the characteristic  radius at a wavelength of 12~$\mu$m for the properties adopted for model (i) is r$\sim 2 \times 10^{12}$ cm, and only slightly larger for model (ii), so that a negligible  fraction of the fine-structure line emission will be obscured.   Other possibilities include a changing ionization structure, possibly as a result of high ionization  in the inner regions, or a non-constant clumping factor.  The ionization fraction is not expected to vary at such large radii due to the absorption of ionizing photons in the inner wind and the  decreasing recombination rates in the declining-density outer wind regions (`freeze-out') \citep{Barlow88}, but could vary in the inner regions.    Unfortunately, the data quality presented here is not adequate to place stringent limits on these effects, but they do indicate that modest degrees of clumping in the wind improve the match with observations for the simple cases considered here.   As shown by Dessert et al (2000), adoption of a clumpy wind with lower mass loss rate increases  the ionic abundances of  Ne$^+$ and S$^{3+}$ compared to an unclumped wind.  The data presented here support the need for a clumpy wind.

\section{Conclusions}

Spatially-resolved imaging and spectroscopy is presented of forbidden emission in the outer wind of the Wolf-Rayet star in the $\gamma$~Velorum binary. The data confirm that forbidden [{S\,\sc iv}] and [{Ne\,\sc ii}] line emission is prominent at large separations (out to $\sim2^{\prime\prime}$ radius and $\sim500\times$ the binary separation) from the star. The images reveal largely symmetric outer-wind emission from [{S\,\sc iv}] and [{Ne\,\sc ii}], suggesting that spherical symmetry  is a reasonable approximation for the outflow.  Spatially-binned spectra are rich in emission lines, which clearly follow three distinct spatial emission profiles. The He and C lines, and the thermal continuum, all trace the distribution of the PSF indicating that the emission is unresolved, as expected for recombination lines and free-free emission which arise primarily from the dense inner regions of the wind. The [{S\,\sc iv}] and [{Ne\,\sc ii}] emission lines have broader spatial profiles.  The [{S\,\sc iv}] line has a broader spatial emission profile than the [{Ne\,\sc ii}] line; this is attributed to  its low critical density which leads to line formation peaking   at larger radii in the wind.

A simple wind model has been constructed, using spherical shells with a 1/$r^{2}$ dependence, to estimate the flux admitted by the T-ReCS slit.  Two cases were considered: a smooth wind with a mass loss rate of $\dot{M}$ = $5.7 {\times}10^{-5} {M_\odot}$  yr$^{-1} $ and a clumped wind with clumping factor $f= 10$ and a mass loss rate of $\dot{M}$ = $1.5 {\times}10^{-5} {M_\odot}$  yr$^{-1} $.  The gradients and ratios of the [{S\,\sc iv}] and [{Ne\,\sc ii}] line profiles at large radii from the WR are well-reproduced by the models, showing that a spherical wind and 1/$r^2$ density dependence are appropriate, if not unique, solutions.  In the inner regions, neither model  fully reproduces the spatial emission profile of the lines,  but the unclumped wind model overpredicts the line intensities substantially. The clumped wind provides a better fit to the data, providing a better match to both the intensity profiles and the spatial distribution of the [{Ne\,\sc ii}] to [{S\,\sc iv}] line ratio.  The wind structure could be probed by higher spatial resolution observations made under better conditions  and  could be examined further by a similar analysis at mid-IR wavelengths of  [{Ne\,\sc iii}]~15.5$\mu$m and  [{S\,\sc iii}]~18.7 $\mu$m  using the {\em James Webb Space Telescope} in the future. 

\section*{Acknowledgments}

The authors thank  Pete Storey for providing the {\sc INTRAT} recombination programme used in this work, and to a referee for providing helpful comments, which led to improvements to the paper.  This paper is based on observations obtained at the Gemini Observatory, which is operated by the Association of Universities for Research in Astronomy, Inc., under a cooperative agreement 
with the NSF on behalf of the Gemini partnership: the National Science Foundation (United 
States), the Science and Technology Facilities Council (United Kingdom), the 
National Research Council (Canada), CONICYT (Chile), the Australian Research Council (Australia), 
MinistŽrio da Cincia, Tecnologia e Inova‹o (Brazil) 
and Ministerio de Ciencia, Tecnolog'a e Innovaci—n Productiva (Argentina).

\bsp

\label{lastpage}

\end{document}